\documentclass{aa}  
\usepackage{graphicx}
\usepackage{txfonts}
\begin{document}
\title{The radial profiles of the different mass components 
in galaxy clusters\thanks{Based on observations collected at the European 
Southern  Observatory (La Silla, Chile)}}
\author{Andrea Biviano\inst{1} \and Paolo Salucci\inst{2}}
\offprints{A. Biviano, biviano@oats.inaf.it}
\institute{INAF/Osservatorio Astronomico di Trieste, via G. B. Tiepolo
11, I-34131, Trieste, Italy \and International School for Advanced
Studies SISSA/ISAS, via Beirut 2--4, I-34013, Trieste, Italy} 
 \authorrunning{Biviano \& Salucci}
\titlerunning{Cluster mass components}
\date{Received; accepted}

\abstract{}{To derive the mass profiles of the different luminous and
dark components in clusters.}{The cluster mass profile is
determined by using the Jeans equation applied to the projected
phase-space distribution of about 3000 galaxies members of 59 nearby
clusters from the ESO Nearby Abell Cluster Survey. The baryonic and
subhaloes mass components are determined from the galaxies'
luminosity-density profiles through scaling relations between
luminosities and baryonic and dark halo masses. The baryonic mass
component associated with the intra-cluster gas is determined using
X-ray data from {\it ROSAT.}}{The baryon-to-total mass fraction
decreases from a value of $\simeq 0.12$ near the center to $\simeq
0.08$ at the distance of $\simeq 0.15$ virial radii, then it increases
again, to reach a value of $\simeq 0.14$ at the virial
radius. Diffuse, cluster-scale dark matter dominates at all radii,
but its contribution to the total mass content decreases outwards to
the virial radius, where the dark matter in subhaloes may contribute
up to $\simeq 23$\%, and the baryons $\simeq 14$\% of the total
mass. The dark mass and diffuse dark mass profiles are well fit by
both cuspy and cored models. The subhalo mass distribution is not
fit by either model.}{} \keywords{cosmology: observations --
galaxies:clusters: general -- galaxies:kinematics and dynamics -- dark
matter}

\maketitle

\section{Introduction}
\label{s-intro}

\begin{verbatim}
  
\end{verbatim}

With the increasing accuracy of cosmological numerical simulations,
the study of the mass profiles of galaxies and galaxy clusters has
become a powerful way to constrain cosmological models. With
numerical simulations, Navarro et al. (1996, NFW hereafter) found that
dark matter (DM hereafter) haloes are characterized by a universal
mass-density profile, simply summarized by two power-law regimes, an
inner one with exponent $-1$, and an outer one with exponent
$-3$. The universality of the profile and the existence of a central
cusp have been confirmed (Moore et al. 1999; Diemand et al. 2004;
Navarro et al. 2004; but see Ricotti 2003).

It is obvious that observational knowledge of the mass
profiles of the dark and baryonic matter in clusters provides crucial
insights into their formation and evolution (see, e.g.,
Gao et al. 2004; Springel et al. 2001). In particular, one aspect of
the $\Lambda$CDM theory that can be tested is the presence of a cusp
in the center of the DM halo. It is well known that on galactic scales
mass profiles of the NFW form are unable to account for the rotation
curves of low-surface brightness galaxies (de Blok \& Bosma 2002),
normal spirals (de Blok et al. 2003; Gentile et al. 2004), and
the fundamental plane of ellipticals (Borriello et al. 2003).

On cluster scales, the situation is far more open. Cluster mass
profiles have been obtained from the analyses of the X-ray emitting
intra-cluster (IC hereafter) gas, of the projected phase-space
distribution of cluster galaxies, and of the gravitational lensing
shear pattern of background galaxies. Most results indicate
consistency with the NFW profile (see, e.g., Allen et al. 2000;
Athreya et al. 2002; Biviano \& Girardi 2003; Jee et al. 2005; Katgert
et al. 2004, hereafter KBM; \L okas et al. 2006; Pratt \& Arnaud 2005;
Rines et al. 2003; Rines \& Diaferio 2006). In some cases a flatter
than NFW profile is preferred (Broadhurst et al. 2005; Ettori et
al. 2002; Kelson et al. 2002; Nevalainen et al. 2000), or even
required (Arieli \& Rephaeli 2003; Demarco et al. 2003; Sand et
al. 2004). While isothermal profiles were rejected by some dynamical
analyses (e.g. Rines et al. 2003; Broadhurst et al. 2005), cored
profiles generally were not excluded (see however Dahle et al. 2003),
as far as the core of the matter distribution is small (Arieli \&
Rephaeli 2003; Biviano \& Girardi 2003; KBM).

In order to allow a proper comparison of the results of simulations
with observations, it is important to determine the {\em total} mass
distribution and its major components, baryons and the subhaloes. In
particular, the baryonic contribution has been shown to be relevant in
clusters, both near the centre because of the substantial contribution
from the cD (e.g. Sand et al. 2004), and in the outer regions, because
of the increasing mass fraction of the IC gas (e.g. \L okas \& Mamon
2003). Recently, KBM have derived a synthetic mass profile of rich
galaxy clusters, using the ESO Nearby Abell Cluster Survey (ENACS)
data-set (Katgert et al. 1998).  KBM found that the total cluster mass
profile is well fitted both by a NFW profile, and by a Burkert (1995)
profile. KBM also found that the total mass profile is very well
traced by the luminosity profile of the early-type galaxies, i.e. the
mass-to-light ratio is almost flat, when only early-type galaxies are
selected, and the brightest members are excluded. The aim of this
paper is to derive the mass distribution of the DM, of the baryonic
matter and of the cluster sub-haloes, separately. We use the same
data-set of KBM analysed in a slightly different way (see
\S~\ref{s-mtot}). We use the X-ray data from Reiprich \& B\"ohringer
(2002) to determine the radial profiles of the IC gas, and the
luminosity-density profiles of cluster galaxies to determine the mass
distributions of the galaxy baryons and of the DM subhaloes.

The structure of this paper is as follows: in \S~\ref{s-mprofs} we
determine the mass profiles of the different cluster components;
in \S~\ref{s-model} we fit models to the observed mass profiles;
in \S~\ref{s-disc} we summarize our results and draw our conclusions.
Throughout this paper we adopt $\rm{H}_0=70$ $\rm{km} \; \rm{s}^{-1}
\; \rm{Mpc}^{-1}$, $\Omega_{m}=0.3$ and $\Omega_{\Lambda}=0.7$.

\section{Mass profiles}
\label{s-mprofs}
\subsection{Total mass}
\label{s-mtot}
We use the data-set of 59 clusters used by KBM and described in
Biviano et al. (2002). These clusters are combined into a single
'ensemble' cluster, in order to improve the rather poor number
statistics of individual clusters. This gives a total sample of $\sim
2900$ member galaxies with positions and redshifts. Note that Sanchis
et al. (2004) have shown that a stacked sample of several
galaxy clusters can be used to determine a reliable average mass
profile of individual clusters.

The stacking is done in the space of normalized clustercentric
distances, $R/r_{100}$, and normalized velocities with respect to the
cluster mean velocity, $(v-\overline{v})/\sigma_v$, where $r_{100}$ is
the radius of the sphere around the cluster centre with mean density
equal to 100 times the critical density, and $\sigma_v$ is the global
velocity dispersion. For the scaling of clustercentric distances,
Biviano et al. (2002) used Carlberg et al.'s (1997) proxy for
$r_{200}$ (the radius of the sphere around the cluster centre with
mean density equal to 200 times the critical density), which is based
on the assumption of an isothermal mass profile. However, the cluster
mass profile determined by KBM is not isothermal, therefore, for the
estimation of both $r_{100}$ and $r_{200}$ (needed for the
determination of the IC gas mass profile, see \S~\ref{s-mgas}), we
prefer to use Popesso et al.'s (2005) relation (see eq.~1 in that
paper) which makes explicit use of the shape of the mass density
profile determined by KBM. The average values of $\sigma_v$ and
$r_{100}$ for our cluster sample are 699 km~s$^{-1}$ and 2.25~Mpc,
respectively, and the average virial mass is $M(<r_{100})=6.5 \,
10^{14} \, M_{\odot}$.
 
\begin{figure}
\begin{center}
\begin{minipage}{0.5\textwidth}
\resizebox{\hsize}{!}{\includegraphics{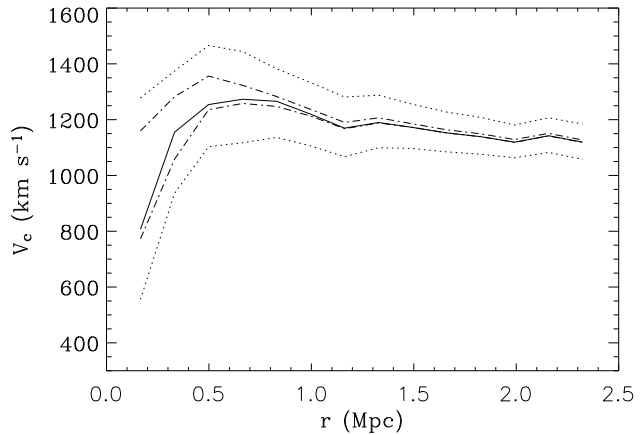}}
\end{minipage}
\end{center}
\caption{The circular velocity profiles, $V_c \equiv (G
M_{tot}/r)^{0.5}$, of the total mass, obtained assuming isotropy
(${\cal A}=0$, solid line), mild radial anisotropy (${\cal A}=0.1$,
lower dash-dotted line), and mild tangential anisotropy (${\cal
A}=-0.6$, upper dash-dotted line). The dotted lines delimit the $\pm
1$-$\sigma$ intervals accounting for both the random errors (as
obtained from 64 bootstrap resamplings of the data in the isotropic
case), and systematic errors (arising from the uncertainty on the
value of the velocity anisotropy ${\cal A}$).}
\label{f-mtot}
\end{figure}

To determine the mass profile of the ensemble cluster ($M_{tot}(r)$ in
the following), we first apply the isotropic Jeans equation on
the early-type cluster members, as described in KBM. The errors
on $M_{tot}(r)$ are obtained from 64 bootstrap resamplings of the data
(see KBM for more details). Near-isotropy for the early-type cluster
members was inferred by KBM from the analysis of the whole velocity
distribution of these galaxies. Support for the isotropic assumption
also comes from the analysis of Biviano \& Katgert (2004) of the
orbits of different cluster galaxy populations. Here we go
beyond KBM's isotropic assumption, and we also consider solutions with
constant orbital anisotropy ${\cal A} \neq 0$, where ${\cal A} \equiv
1-\rm{<}v_t^2 \rm{>}/\rm{<} v_r^2\rm{>}$, and $\rm{<} v_r^2 \rm{>}$,
$\rm{<} v_t^2 \rm{>}$ are the mean squared components of the radial
and tangential velocity, respectively (see, e.g., Binney \& Tremaine
1987). Since KBM constrained the velocity anisotropy of the early-type
galaxy population to lie in the range $-0.6 \la {\cal A} \la 0.1$, we
determine $M_{tot}(r)$ for the two extreme cases ${\cal A}=-0.6$ and
${\cal A}=0.1$, using the anisotropic Abel inversion (Mamon \& Bou\'e
2006) and Jeans equations. The results are shown in
Fig.~\ref{f-mtot}, where we display the circular velocity profiles,
$V_c \equiv (G M_{tot}/r)^{0.5}$. While the isotropic
solution for the total mass profile is not significantly different
from that derived by KBM, orbital anisotropy
has some effect on the uncertainty of the total mass profile.

\subsection{Baryonic mass: IC gas}
\label{s-mgas}
To determine the mass profile of the IC gas ($M_{gas}(r)$ in the
following), we use the IC gas density profiles. Unfortunately these
are available only for a subset of our 59 cluster sample. This
subset is characterized by a larger average velocity dispersion than
the whole sample. Since the shape of the IC gas density profile
depends on the cluster X-ray temperature (Mohr et al. 1999), and hence
also on the cluster velocity dispersion, we cannot use the subset of
clusters with available X-ray data as representative of the whole
sample.

\begin{figure}
\begin{center}
\begin{minipage}{0.5\textwidth}
\resizebox{\hsize}{!}{\includegraphics{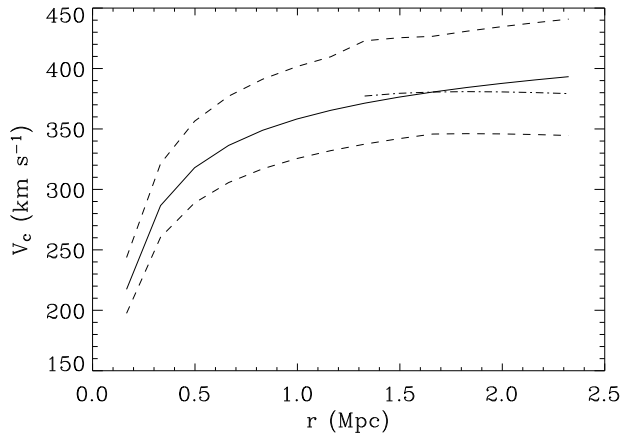}}
\end{minipage}
\end{center}
\caption{The contribution of the IC gas mass to the circular
velocity profile, $V_c \equiv (G M_{gas}/r)^{0.5}$ (solid line),
obtained from the sample of Reiprich \& B\"ohringer (2002). The
dash-dotted line shows the effect of using Neumann's (2005) steeper IC
gas density profile in the radial range $>r_{500}$. The dashed lines
delimit the $\pm 1$-$\sigma$ intervals accounting for the
uncertainties in the best-fit parameters of the IC gas density
profile, as well as for the uncertainty in the value
$(M_{gas}/M_{tot})(r_{200})$.}
\label{f-mgas}
\end{figure}

We proceed instead as follows. We consider Reiprich \&
B\"ohringer's (2002) sample of clusters for which
the best-fit parameters of the $\beta$-profiles
(Cavaliere \& Fusco-Femiano, 1978) are available,

\begin{equation}
\rho_{IC} = \rho_0 [1+(r/r_c)^2]^{-(3/2)\beta},
\end{equation}

We then consider the distribution of velocity
dispersions of our 59 clusters, and convert it into a distribution of
pseudo X-ray temperatures, using the empirical relation of Girardi et
al. (1996). We generate 500 bootstrap sets of this pseudo-$T_X$
distribution. For each of these bootstrap sets, we then extract 59
clusters from the sample of Reiprich \& B\"ohringer (2002), chosen to
have a $T_X$-distribution as close as possible to the pseudo-$T_X$
distribution of the bootstrap set. We finally compute the average
values of $r_c$ and $\beta$ for the $59 \times 500$ extracted
clusters (many of the clusters are of course extracted more than once,
as expected for a bootstrap procedure), $<\beta>=0.625 \pm 0.007$,
and $<r_c>=(0.0636 \pm 0.0004) \, r_{200}$. We use these values to
define the average gas density profile, which is meant to be
representative of our cluster sample. Integration of this average gas
density profile provides the IC gas mass profile, apart from a
constant, that we fix to the average gas-to-total mass fraction at
$r_{200}$ as determined by Ettori (2003) for a sample of nearby
clusters, $0.11_{-0.02}^{+0.03}$.
 
There is an additional uncertainty that is related to the
extrapolation of the observed IC gas profiles to $r_{200}$.  Most of
the IC gas profiles on which $M_{gas}(r)$ is based are 
determined from data at radii smaller than $\sim r_{500}$ (the radius
of the sphere around the cluster centre with mean density equal to 500
times the critical density) corresponding to $\sim 1$~Mpc in our
cluster sample. In order to deal with the systematic uncertainty
related to this extrapolation, we take into consideration the 
analysis of Neumann (2005). Neumann has analysed a sample of
14 nearby clusters, for which she has been able to trace the X-ray
emission beyond $r_{200}$. She concluded that the IC gas density
profile steepens with increasing radius. Out to 1.2 $r_{200}$ Neumann
found a best-fit $\beta$-profile\footnote{Among the subsamples of
Neumann (2005) we have chosen her no.1, based on the similarity
between the $T_X$ distribution of this subsample and of our clusters.}
with $r_c=0.19 \, r_{200}$ and $\beta=0.8$.  The fitted profile is
therefore significantly steeper than the mean profile we have found by
using the sample of Reiprich \& B\"ohringer (2002).

We then adopt Neumann's best-fit to estimate the
uncertainty involved in the extrapolation of our average IC gas
density profile to radii $> r_{500}$. We find that $M_{gas}$ is not
strongly modified beyond $r_{500}$ (and out to $1.5 \, r_{200}$), 
despite the substantial change in the IC density profile, since the
normalisation of the gas mass profile at $r_{200}$ is fixed,
$(M_{gas}/M_{tot})(r_{200})=0.11_{-0.02}^{+0.03}$. Hence, the
uncertainty on the resulting $M_{gas}$ is mostly driven by the
uncertainty in the normalization of the gas-to-total mass fraction at
$r_{200}$. In Fig.~\ref{f-mgas} we show the resulting contribution of
the IC gas mass component to the circular velocity profile, with its
confidence interval, accounting for the uncertainties in the best-fit
parameters of the IC gas density profile, and in the normalisation
value $(M_{gas}/M_{tot})(r_{200})$.

\subsection{Baryonic mass: galaxies}
\label{s-mgals}
In order to determine the baryonic mass profile in the galactic
component ($M_{gal}^{lum}(r)$ hereafter) we consider separately the
two classes of early- and late-types (we use the data-set of Thomas \&
Katgert 2006 for the morphological/spectral classification of ENACS
galaxies). The baryonic mass profiles of these two galaxy classes are
obtained following the prescriptions of KBM for the derivation of
their number density profiles (see Appendix B in KBM; see also Biviano
\& Katgert 2004). KBM's methodology accounts for both the incomplete
azimuthal coverage of the ENACS observations (Katgert et al. 1996,
1998), and the fact that different clusters are sampled out to
different fractions of their virial radii.

In KBM the number density and the luminosity density profiles were
derived; here we proceed further by assuming (for each galaxy class) a
relation converting a given galaxy luminosity to its baryonic
component. For the early class we take Borriello et al.'s (2003; see
their eq.15) relation, and convert their magnitudes to the $R$-band
magnitudes of ENACS galaxies (Katgert et al. 1996), using the
relations of Fukugita et al. (1995). For late-type galaxies the
relationship between luminosities and baryonic masses is
taken from Salucci \& Persic (1999; see their eqs.~3--5).
 
Since the ENACS sample is not complete, we need to correct the density
profiles for incompleteness. We estimate that the ENACS sample is
roughly 75\% complete down to an apparent magnitude $R=16.5$ (see
Fig.~4 in Katgert et al. 1998), and then the completeness rapidly
drops for fainter magnitudes. At the median redshift of our clusters
($z=0.064$), $R=16.5$ corresponds to an absolute magnitude
$M_R=-20.8$.  For simplicity, we then assume 0.75 completeness down to
$M_R=-20.8$, and zero at fainter magnitudes. 

In order to derive the luminosity of galaxies fainter than
$M_R=-20.8$, we use the $R$-band luminosity function of Lugger (1986),
that is a Schechter (1976) luminosity function with
$M_R^{\star}=-21.9$ and $\alpha=-1.24$. We integrate it between
$M_R=-20.8$ and the magnitude corresponding to $0.01 L_R^{\star}$,
where $L_R^{\star}$ is the luminosity corresponding to
$M_R^{\star}$. We further assume that most faint cluster galaxies with
$M_R \geq -20.8$ are dwarf spheroidals, i.e. early-type galaxies.  We
find that galaxies fainter than $M_R=-20.8$ contribute 25\% of the
galactic baryonic mass in a cluster. We then correct the observed
baryonic profiles for the faint galaxies contribution, and for the
additional factor $1.33=1/0.75$, that accounts for the overall
spectroscopic incompleteness.

\begin{figure}
\begin{center}
\begin{minipage}{0.5\textwidth}
\resizebox{\hsize}{!}{\includegraphics{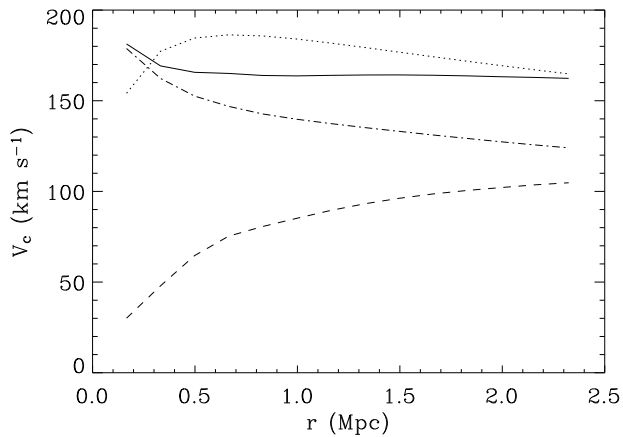}}
\end{minipage}
\end{center}
\caption{The contributions to the circular velocity profile of
the baryonic mass in all cluster galaxies (solid line), in early-type
cluster galaxies (dash-dotted line), and in late-type galaxies (dashed
line). The dotted line shows the profile derived by \L okas \& Mamon
(2003) for the Coma cluster, after appropriate scaling (see text).}
\label{f-mgals}
\end{figure}

The resulting contribution of the baryonic mass in cluster
galaxies to the circular velocity profile is shown in
Fig.~\ref{f-mgals}. In the same figure we also show the separate
contributions of the baryonic mass in early-type and
late-type galaxies. The profile of the baryonic mass in
early-type galaxies is similar to that of the total mass, but more
centrally concentrated. The profile of late-type galaxies is instead
less centrally concentrated than the total mass profile.  Early-type
galaxies dominate the galactic baryonic budget within the virial
radius, but the contribution of late-type galaxies to this budget
increases with radius, from only 3\% near the center to $\sim 40$\%
within the virial radius.

In order to put our galaxy baryonic mass profile in context, we
compare it with that derived by \L okas \& Mamon (2003) for the Coma
galaxy cluster. First we compute the ratios between the Coma virial
radius and total mass and the corresponding average values for our 59
clusters (these ratios are 1.3 and 1.9, respectively). We then rescale
\L okas \& Mamon's profile using these scaling factors, and assuming
that the cluster baryonic mass scales proportionally to the cluster
total mass.  The result is shown as a dashed line in
Fig.~\ref{f-mgals}. Clearly, \L okas \& Mamon's profile is less
concentrated than ours, but it does not differ by more than $\pm 20$\%
over the whole radial range considered here. We consider this a
remarkable agreement, given the fact that Coma is not a typical galaxy
cluster. While 80\% of our clusters are dominated by a single brightest
cluster galaxy (BCG hereafter), two BCGs are present in the inner
region of Coma, which is in fact substructured (see, e.g., Adami et
al. 2005 and references therein). The presence of substructures in the
Coma cluster core probably reduces the concentration of its baryonic
mass profile.

\subsection{Dark mass: subhaloes}
\label{s-mhaloes}
The distinction between DM in subhaloes and diffuse DM is to some
extent a matter of definition. Here we consider as subhaloes all
visible cluster galaxies (including those we miss because of the
completeness limit of the ENACS survey), with the exception of the
BCG. Since the BCG is centrally located, its diffuse stellar and DM
haloes are generally considered part of the diffuse IC material (Lin
\& Mohr 2004; Murante et al. 2004).

We compute the subhalo mass profile ($M_{sub}(r)$ in the following)
from the average luminosity density profiles of the cluster member
galaxies, by adopting a Hubble-type dependent scaling relation between
a galaxy luminosity and its halo mass (see Shankar et al. 2006). As in
\S~\ref{s-mgals}, we apply the needed corrections for
incompleteness. Galaxies fainter than the spectroscopy limit of the
ENACS survey ($M_R \simeq -20.8$) contribute roughly 30\% of the
subhaloes mass. An additional factor of 1.33 must be included to account
for the average incompleteness of the ENACS spectroscopic sample (see
\S~\ref{s-mgals}).

\begin{figure}
\begin{center}
\begin{minipage}{0.5\textwidth}
\resizebox{\hsize}{!}{\includegraphics{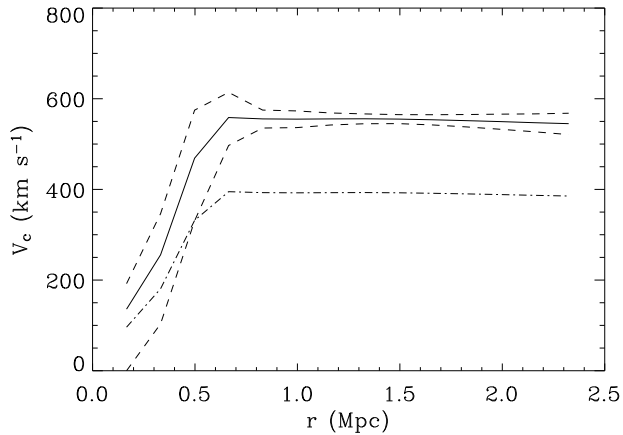}}
\end{minipage}
\end{center}
\caption{The contribution of the mass in subhaloes to the
circular velocity profile, $V_c \equiv (G M_{sub}/r)^{0.5}$ (solid
line) and its 1-$\sigma$ confidence interval (dashed lines) accounting
for both the random and the systematic errors (the latter being
related to the uncertainty in the size of the central stripping
region).  The dash-dotted line corresponds to the scenario in which
subhaloes lose 50\% of their total mass.}
\label{f-msubh}
\end{figure}

The resulting subhalo mass profile needs however to be modified to
take into account the fact that, unlike in the field, galaxies in
clusters are so densely packed that their haloes would overlap. This
overlap does not in fact occur because their haloes are tidally
stripped as they pass through dense regions. Numerical simulations
predict a flattening of the density profiles of subhaloes near the
cluster centre, and as much as a 50\% loss of the total mass in
subhaloes, due to the stripping process (see, e.g., Gao et
al. 2004). Gravitational lensing observations confirm that the haloes
of galaxies near the cluster centres are indeed less massive than
those of field galaxies (Natarajan et al.  2002; Gavazzi et al. 2004).

In order to take into account the effects of tidal stripping near the
cluster centre, and given the current, rather loose, observational
constraints on this topic, we try the following simplified approach.
We modify our estimate of the subhalo mass as follows. First, we
assume that subhalo stripping only occurs within a critical radius
$R_{BCG}$, and that stripping is more effective as the clustercentric
radius decreases. In practice, we assume $M_{sub} \approx 0$ at the
cluster centre, and smoothly interpolate to the unstripped $M_{sub}$
at the radius $R_{BCG}$.

We consider it reasonable to identify the value of $R_{BCG}$
with the radius of the halo of the BCG. BCGs can be followed
photometrically out to $\sim 0.4$ Mpc (see, e.g., Gonzalez et
al. 2005). The haloes of BCGs have been modelled with a Burkert
profile with scale radius $r_s=(2.6 \pm 0.2) R_e$ (Borriello et
al. 2003), where $R_e=18 \pm 2$ kpc (Nelson et al. 2002), implying
that 1/2 of the total BCG mass is contained in a sphere of $\sim 0.5$
Mpc radius. Moreover, numerical simulations predict that the density
profile of subhaloes deviates from the overall density profile of DM
inward of $\sim 0.4 \, r_{200} \simeq 0.7$ Mpc (see Fig.~1 in Gao et
al. 2004). Therefore we take $R_{BCG} \sim 0.5 \pm 0.2$ Mpc. The
uncertainty in the value of $R_{BCG}$ increases the error of $M_{sub}$
in the central region.

The resulting contribution of $M_{sub}$ to the circular velocity
profile is shown in Fig.~\ref{f-msubh}. We also display the circular
velocity profile corresponding to the rather extreme scenario in which
50\% of the total subhaloes mass is lost at any radii ({\em strong
stripping} scenario hereafter.)

\begin{figure*}
\begin{center}
\begin{minipage}{0.95\textwidth}
\resizebox{\hsize}{!}{\includegraphics{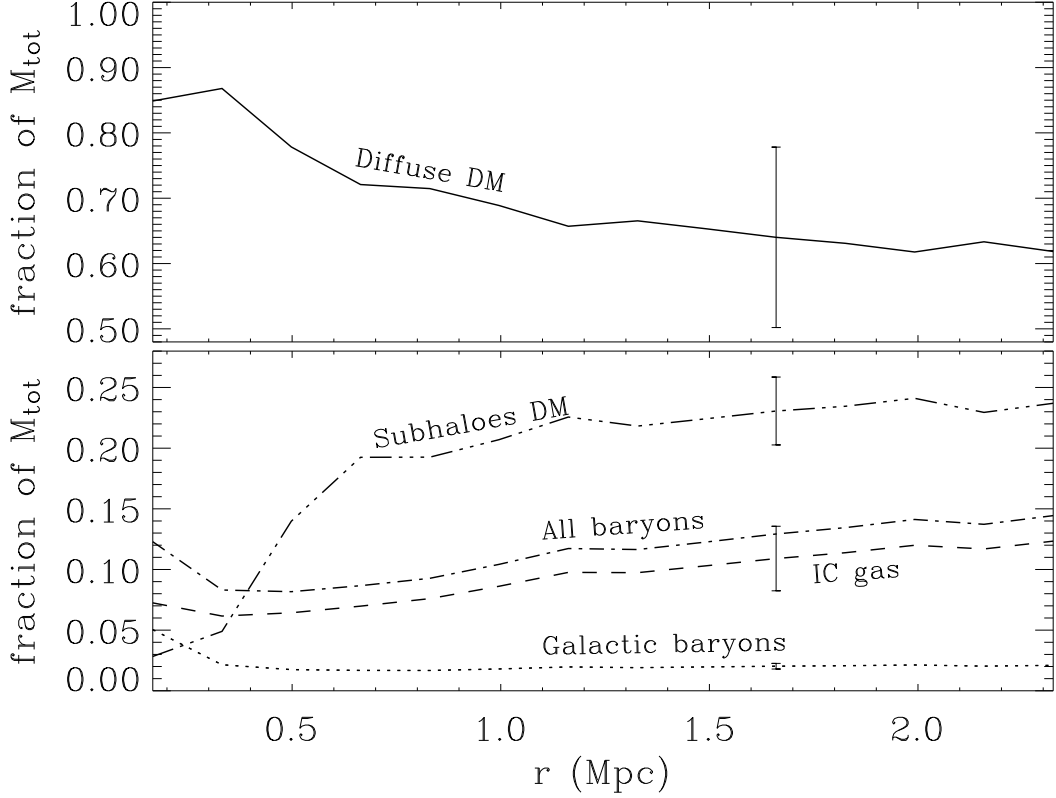}}
\end{minipage}
\end{center}
\caption{The ratios between the mass profiles of the different
cluster components and the total mass profile, within a given
radius. Solid line: $(M_{dark}^{diff}/M_{tot})(r)$; dotted line:
$(M_{gal}^{lum}/M_{tot})(r)$; dashed line: $(M_{gas}/M_{tot})(r)$;
dot-dashed line: $[(M_{gal}^{lum}+M_{gas})/M_{tot}](r)$; triple
dot-dashed line: $(M_{sub}/M_{tot})(r)$. The vertical lines indicate
$\pm 1$-$\sigma$ error bars at the $r_{200}$ radius. The error bar on
$(M_{gal}^{lum}/M_{tot})(r)$ is very small and not visible in this
plot. The error bar on $[(M_{gal}^{lum}+M_{gas})/M_{tot}](r)$ is not
shown, since it is almost identical to the error bar on
$(M_{gas}/M_{tot})(r)$.}
\label{f-fract}
\end{figure*}

\subsection{Dark mass: diffuse}
\label{s-mdark}
The diffuse DM profile is obtained from the total mass profile
by subtracting from it the baryonic mass profile, $M_{dark}(r) \equiv
M_{tot}(r)-M_{gal}^{lum}(r)-M_{gas}(r)$.

Baryons are a minor though not irrelevant component of the total
cluster mass. On the other hand, a non-negligible fraction of the
total cluster mass is in subhaloes. When both the baryonic and
subhalo masses are subtracted from the total mass, we obtain what we
call the diffuse DM profile $M_{dark}^{diff}(r) \equiv
M_{tot}(r)-M_{gal}^{lum}(r)-M_{gas}(r)-M_{sub}(r)$.
 
\subsection{Relative fractions}
\label{s-fract}
\begin{table*}
\begin{center}
\begin{tabular}{lllll}
\hline
Mass      & \multicolumn{4}{c}{fraction of total mass within} \\
component & $0.1 \, r_{100}$ & $r_{500}$ & $r_{200}$ & $r_{100}$ \\
\hline
& & & & \\
$M_{gas}$  & $0.07 \pm 0.03$  & $0.09 \pm 0.03$          & $0.11 \pm 0.03$          & $0.12 \pm 0.03$ \\
& & & & \\
$M_{gal}^{lum}$  & $0.04 \pm 0.01$  & $0.019 \pm 0.003$        & $0.020 \pm 0.002$        & $0.021 \pm 0.002$ \\
& & & & \\
$M_{sub}$  & $0.04 (0.02) \pm 0.04$  & $0.22 (0.11) \pm 0.04$   & $0.23 (0.12) \pm 0.03$   & $0.23 (0.12) \pm 0.03$ \\
& & & & \\
$M_{dark}^{diff}$ & $0.85 (0.86) \pm 0.41$  & $0.67 (0.78) \pm 0.20$   & $0.64 (0.75) \pm 0.14$   & $0.63 (0.74) \pm 0.13$ \\
& & & & \\
\hline
\end{tabular}
\caption{Relative contributions of the different cluster mass
components at four characteristic radii. Values in brackets are for
the strong stripping scenario.}
\label{t-fract}
\end{center}
\end{table*}

The mass fractions of the different cluster components relative to the
total cluster mass are shown in Fig.~\ref{f-fract} as {\em cumulative}
mass fractions within the radii on the x-axis.  For
clarity, confidence intervals are not shown. We only show the
error-bars at $r_{200}$. Error intervals at other
characteristic radii are listed in Table~\ref{t-fract} (see also
Figs.~\ref{f-mtot}--\ref{f-msubh}). The largest uncertainty is on
$M_{tot}$, and this affects $M_{dark}^{diff}$ (see also
Fig.~\ref{f-vcall}).

Among the baryonic components, the IC gas is clearly dominant at all
radii. However, very near the cluster center, galaxies contribute
almost as much baryonic mass as the gas.  This is due to the presence
of the cD and/or very bright galaxies near the cluster center (the
phenomenon also known as 'luminosity segregation', see, e.g., Biviano
et al. 1992, 2002). Baryons in galaxies and the IC gas are more,
respectively less, centrally concentrated than the total mass; hence,
the ratio of the total baryonic mass to the total
cluster mass has a minimum at $\sim 0.15 \, r_{100} \simeq 0.3$--0.4
Mpc.  The ratio between the baryonic and total mass profiles
is constant to within $\pm 30$\% out to the cluster virial radius.

The mass contribution from the subhaloes is small near the
center, where stripping may occur (see \S~\ref{s-mhaloes}), and then
increases out to 1.1 Mpc, when it reaches an approximately constant
fraction of $\sim 0.2$--0.25.

Diffuse DM is the dominant mass component at all radii. Its
contribution to the total mass almost monotonically decreases with
radius, except perhaps at the center, where the galactic baryons
contribute significantly.

The relative contributions of diffuse and subhaloes DM change if one
considers the strong stripping scenario.  In this scenario, the
contribution of subhaloes to the total mass budget is similar to that
of all baryons (see the values listed in brackets in
Table~\ref{t-fract}) and diffuse DM contributes 3/4 of the total mass.

\section{Model fits}
\label{s-model}
We fit the observed DM profiles with two models, the NFW and Burkert
profiles. Both models are characterized by the same asymptotic slope
at large radii, $\rho(r) \propto r^{-3}$, but the former is
characterized by an inner cusp, the latter by a core. By comparing the
results of the two model fits we can address the debated issue of the
reality of the inner cusp that numerical simulations predict to exist
in DM haloes.

The NFW mass density profile model can be written as:
\begin{equation}
\rho_{NFW}=\frac{\rho_0}{(r/r_s)(1+r/r_s)^2}
\end{equation}
and the Burkert profile as:
\begin{equation}
\rho_{Burkert}=\frac{\rho_0}{(1+r/r_0)[1+(r/r_0)^2]}.
\end{equation}
They are both characterized by a scale radius, $r_s$ or $r_0$, which,
in the case of the NFW profile is usually referred to as the inverse
of the concentration parameter, $c \equiv r_{100}/r_s$, and, in
the case of the Burkert profile, corresponds to the radius at which
the central density drops by a factor 4.

\begin{figure*}
\begin{center}
\begin{minipage}{0.95\textwidth}
\resizebox{\hsize}{!}{\includegraphics{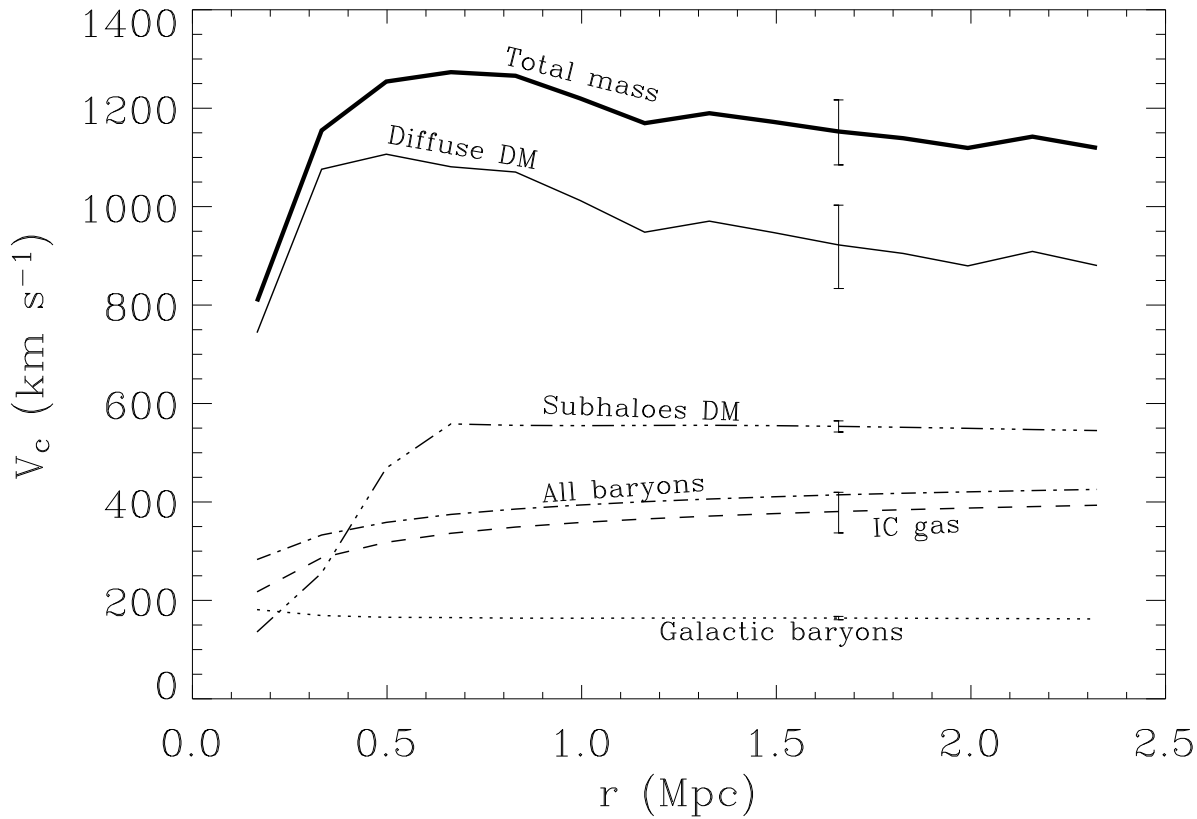}}
\end{minipage}
\end{center}
\caption{The contributions of the different cluster mass components to
the circular velocity profile, $V_c \equiv (G M/r)^{0.5}$. Upper
(thick) solid line: $M_{tot}$; lower (thin) solid line:
$M_{dark}^{diff}$; dotted line: $M_{gal}^{lum}$; dashed line:
$M_{gas}$; dot-dashed line: $M_{gal}^{lum}+M_{gas}$; triple dot-dashed
line: $M_{sub}$. For clarity, only representative
1-$\sigma$ error bars (at the $r_{200}$ radius) are displayed. Note
that the error bar on the galaxies' baryonic profile is very
small and not visible in this plot. The error bar on the total
baryonic profile is not shown, since it is almost identical
to the error on the IC gas baryonic profile.}
\label{f-vcall}
\end{figure*}

\begin{figure*}
\begin{center}
\begin{minipage}{0.95\textwidth}
\resizebox{\hsize}{!}{\includegraphics{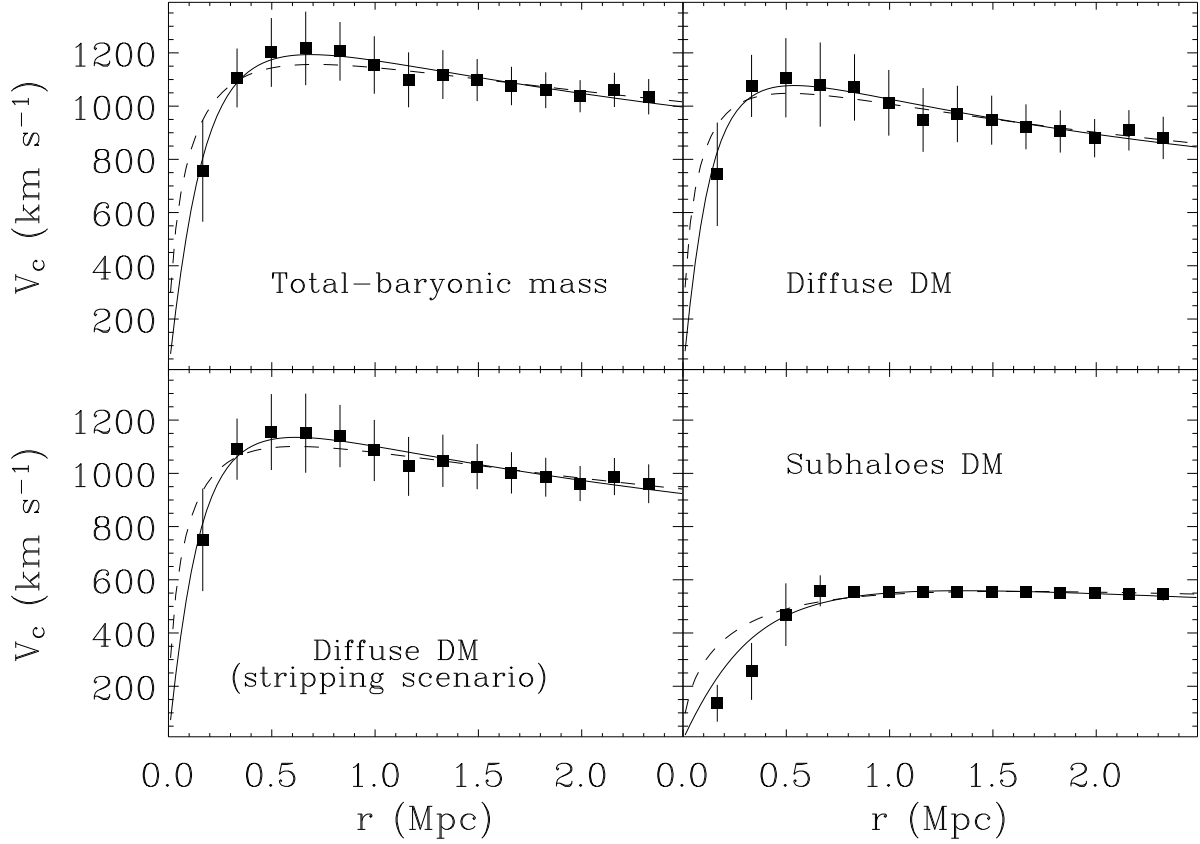}}
\end{minipage}
\end{center}
\caption{Top-left panel: The circular velocity profile, $V_c \equiv (G
M_{dark}/r)^{0.5}$, corresponding to the cluster DM profile (filled
squares with 1-$\sigma$ error bars), and the best-fitting Burkert (solid
line) and NFW (dashed line) profiles. Top-right panel: same as
top-left panel, for the circular velocity profile corresponding to
$M_{dark}^{diff}$. Bottom-left panel: same as top-left panel, for the
circular velocity profile corresponding to the cluster diffuse DM
profile, assuming tidal stripping of 50\% of the subhaloes
mass. Bottom-right panel: same as top-left panel, for the circular
velocity profile corresponding to $M_{sub}$.}
\label{f-fits}
\end{figure*}

In order to determine the best fit parameters of the two profiles, we
fit the observed circular velocity profiles, $V_c$.
These are displayed in Fig.~\ref{f-vcall} for the different mass
components. In order to determine the quality of the fit, we make use
of the standard $\chi^2$ analysis, applied on the mass density
profiles, $\rho(r)$, instead of on the circular velocity profiles,
$V_c(r)$, since the error bars of $\rho(r)$ at different radii are
independent of one another,

We start by considering the circular velocity profile corresponding to
$M_{dark}(r)$.  The best-fit NFW and Burkert profiles are displayed in
Fig.~\ref{f-fits}, top-left panel.  The results are given in
Table~\ref{t-fits}, where the characteristic radii are all given in
units of $r_{100}$ ($\simeq 2.25$ Mpc). Both the NFW and Burkert
models provide acceptable fits to the circular velocity profile
corresponding to $M_{dark}(r)$. The best-fit NFW profile has a a
concentration parameter $c=7 \pm 1$, which is in line with the
predictions of cosmological simulations ($c=8.3$ for clusters of
$M(<r_{100})=6.5 \, 10^{14} \, M_{\odot}$, see Dolag et al. 2004
\footnote{Dolag et al.'s (2004) definition of $c$ is different from
that used here; we converted their values to make them consistent
with the definition of $c$ used in the present paper.}).

To fit the circular velocity profile corresponding to the diffuse DM
component, both the NFW and the Burkert mass distributions become more
concentrated (see Fig.~\ref{f-fits}, top-right panel, and
Table~\ref{t-fits}). This is because the subtracted subhalo mass
distribution is less concentrated than the total DM distribution. The
best-fit $r_0$ value of the Burkert profile is small, but still
significantly different from zero. The best-fit NFW profile has a
concentration $c=10_{-2}^{+4}$.  The value of
concentration does not change significantly, if we consider the
strong stripping scenario ($c=8 \pm 2$, see Fig.~\ref{f-fits},
bottom-left panel, and values in brackets in Table~\ref{t-fits}).

\begin{table}
\begin{tabular}{cccc}
\hline
Mass profile & Model & $r_s$ or $r_0$   & reduced $\chi^2$ \\
             &       & ($r_{100}$ units) &  \\
\hline
& & & \\
$M_{dark}$ & NFW & $0.15 \pm 0.03$ & 0.3 \\
           & Burkert & $0.09 \pm 0.01$ & 0.4 \\
& & & \\
$M_{dark}^{diff}$ & NFW & $0.10 (0.13) \pm 0.03$ & 0.3 (0.2) \\
           & Burkert & $0.07 (0.08) \pm 0.02$ & 0.3 (0.3)\\
& & & \\
$M_{sub}$  & NFW & $0.33 \pm 0.05$ & 2.6 \\
           & Burkert & $0.19 \pm 0.02$ & 2.4 \\
& & & \\
\hline
\end{tabular}
\caption{Results of model fits to the observed mass profiles. Values
in brackets are for the case that 50\% of the total mass in subhaloes is
tidally stripped.}
\label{t-fits}
\end{table}

Neither the NFW nor the Burkert model provides acceptable fits to the
circular velocity profile corresponding to $M_{sub}(r)$ (see
Fig.~\ref{f-fits}, bottom-right panel, and Table~\ref{t-fits}). The
rather large best-fit values of $r_s$ and $r_0$ (for the NFW and
Burkert model, respectively) reflect the fact that subhaloes 
tend to avoid the central cluster region (this was already
quite clear from Fig.~\ref{f-fract}).

\section{Summary and conclusions}
\label{s-disc}

We have obtained the mass profiles of the different cluster
components, namely the baryons (in galaxies and the IC gas), the
subhaloes (galactic halo DM), and the diffuse DM, out to the cluster
virial radius, $r_{100} \simeq 2.25$ Mpc.  

We have determined a total budget of the different mass components at
different radii. The diffuse, cluster-scale, DM is the dominant
cluster mass component at all radii. The total baryonic mass fraction,
resulting from the summed contribution of the galactic and IC baryons,
is 14\% of the total cluster mass within the virial radius. The
baryonic mass fraction first decreases, and then increases again with
radius, changing by $\pm 30$\% within the cluster virial radius. The
galaxy baryonic component is always a small amount of the total mass,
except near the cluster center, where galaxies contribute almost as
much baryonic mass as the IC gas (because of the centrally located
cD).  Considering clusters as a cosmic laboratory, the low relative
fraction of baryons in galaxies indicates how the star-formation in
the Universe has been inefficient, since only $\simeq 14$\% of the
baryonic mass content in clusters has been transformed in long lived
stars.

The baryonic mass profile of early-type galaxies has a shape similar
to that of the total mass profile (in agreement with e.g. van der
Marel et al. 2000, KBM, Biviano \& Girardi 2003, \L okas \& Mamon
2003). On the other hand, the baryonic mass profile of late-type
galaxies is less concentrated than the total mass profile.  Most
galaxy baryons in clusters are contributed by early-type galaxies. The
IC gas-to-total mass fraction increases with radius as $r^{0.4}$
beyond $\sim 0.2 \, r_{100}$, in agreement with previous findings
(e.g. Allen et al. 2000, \L okas \& Mamon 2003).

We estimate the subhalo mass fraction, 12--23$\%$, which is in
agreement with the value predicted by the results of cosmological
numerical simulations (see, e.g.  Takahashi et al. 2002; van den Bosh
et al. 2005; Gill et al. 2004), and
estimated observationally by Natarajan et al. (2002).

We obtain good fits to the DM and diffuse DM profiles with the NFW and
Burkert models, while the subhalo mass profile cannot be fitted by
these models. For the DM profile we obtain
\begin{eqnarray}
M_{dark}/(10^{14} \, M_{\odot}) = 4.7 \, [\ln(1+r/0.33)- \nonumber \\ r/0.33 (1+r/0.33)^{-1}]
\end{eqnarray}
with $r$ in Mpc, for the NFW fit, and
\begin{eqnarray}
M_{dark}/(10^{14} \, M_{\odot}) = 1.6 \, \{\ln(1+r/0.21)- \nonumber \\ \arctan(r/0.21) + 0.5 \ln[1+(r/0.21)^2]\}
\end{eqnarray}
for the Burkert fit, for an average cluster of mass 
$M_{tot}(<r_{100})= 6.5 \, 10^{14} \, M_{\odot}$, and $r_{100} = 2.25$~Mpc.
For the diffuse DM profile we obtain 
\begin{eqnarray}
M_{dark}^{diff}/(10^{14} \, M_{\odot}) = 2.8 \, [\ln(1+r/0.24)- \nonumber \\ r/0.24 (1+r/0.24)^{-1}]
\end{eqnarray}
for the NFW fit, and
\begin{eqnarray}
M_{dark}^{diff}/(10^{14} \, M_{\odot}) = 1.3 \, \{\ln(1+r/0.19)- \nonumber \\ \arctan(r/0.19) + 0.5 \ln[1+(r/0.19)^2]\}
\end{eqnarray}
for the Burkert fit. In the strong stripping scenario (see
\S~\ref{s-mhaloes}) we obtain instead
\begin{eqnarray}
M_{dark}^{diff}/(10^{14} \, M_{\odot}) = 3.6 \, [\ln(1+r/0.28)- \nonumber \\ r/0.28 (1+r/0.28)^{-1}]
\end{eqnarray}
for the NFW fit, and
\begin{eqnarray}
M_{dark}^{diff}/(10^{14} \, M_{\odot}) = 1.3 \, \{\ln(1+r/0.19)- \nonumber \\ \arctan(r/0.19) + 0.5 \ln[1+(r/0.19)^2]\}
\end{eqnarray}
for the Burkert fit.

Note that the best-fit NFW concentration values ($c=7 \pm 1$ and
$c=10_{-2}^{+4}$, respectively, for the DM and DM diffuse
profiles) are similar to the predictions of $\Lambda$CDM cosmological
simulations (e.g. Dolag et al. 2004).

Our results are for clusters on average, and it is possible that
individual clusters are characterized by different types of mass
profiles (see, e.g., Ettori et al. 2002).  Our results provide new
constraints relative to most recent analyses (e.g. van der Marel et
al.  2000; KBM) in that we derive the {\em dark}, rather than the {\em
total} mass distribution. \L okas \& Mamon (2003) did consider the
{\em dark} matter distribution, but only of one cluster (Coma), while
we have examined a sample of 59 clusters. Moreover, they did not
subtract the contribution of subhaloes from the DM profile.

The major uncertainty in our analysis is due to the still rather
limited number of available cluster galaxies with redshifts. Other
sources of uncertainties (such as systematics in the knowledge of
the IC gas density profile at large radii, or of the subhalo
stripping efficiency), are smaller.  With a larger data-set it will be
possible to improve the constraints on the cluster mass profile, due
to a better definition of the projected phase-space distribution of
cluster members, and their orbital anisotropy.  Improvements over the
current analysis are therefore expected from the use of larger
samples of clusters extracted from the Sloan Digital Sky Survey
(e.g. Abazajian et al. 2004), which should allow a factor $\sim 5$
increase in size with respect to the ENACS data-set used here (see
Goto 2005).

\begin{acknowledgements}
We thank Gary Mamon for useful discussion. We also thank the anonymous
referee for useful and constructive remarks. This research was
partially supported by the Italian Ministry of Education, University,
and Research (MIUR grant COFIN2001028932 "Clusters and groups of
galaxies, the interplay of dark and baryonic matter")
\end{acknowledgements}


\begin{thebibliography}{}
\bibitem[2004]{aba04} Abazajian, L., et al. 2004, AJ, 128, 502
\bibitem[2005]{ada05} Adami, C., Biviano, A., Durret, F., \& Mazure, A. 2005, A\&A, 443, 17
\bibitem[2000]{all00} Allen, S.W., Ettori, S., \& Fabian, A.C. 2000, MNRAS, 324, 877
\bibitem[2003]{ari03} Arieli, Y., \& Rephaeli, Y. 2003, New A, 8, 517
\bibitem[2002]{ath02} Athreya, R.M., Mellier, Y., van Waerbeke, L., et al. 2002, A\&A, 384, 743
\bibitem[1987]{bin87} Binney, J. \& Tremaine, S. 1987, Galactic Dynamics (Princeton: Princeton University Press)
\bibitem[2003]{biv03} Biviano, A. \& Girardi, M. 2003, ApJ, 585, 205
\bibitem[2002]{biv92} Biviano, A., Girardi, M., Giuricin, G., Mardirossian, F., Mezzetti, M. 1992, ApJ, 396, 35
\bibitem[2002]{biv04} Biviano, A., \& Katgert, P. 2004, A\&A, 424, 779
\bibitem[2002]{biv02} Biviano, A., Katgert, P., Thomas, T., \& Adami, C. 2002, A\&A, 387, 8
\bibitem[2003]{bor03} Borriello, A., Salucci, P., \& Danese, L. 2003, MNRAS, 341, 1109
\bibitem[2005]{bro05} Broadhurst, T., Takada, M., Umetsu, K., et al. 2005, ApJ, 619, L143
\bibitem[1995]{bur95} Burkert, A. 1995, ApJ, 447, L25
\bibitem[1997]{car97} Carlberg, R.G., Yee, H.K.C., \& Ellingson, E. 1997, ApJ, 476, L7
\bibitem[1978]{cav78} Cavaliere, A., \& Fusco-Femiano, R. 1978, A\&A, 70, 677
\bibitem[2003]{dah03} Dahle, H., Hannestad, S., \& Sommer-Larsen, J. 2003, ApJ, 588, L73
\bibitem[2002]{deb02} de Blok, W.J.G., \& Bosma, A. 2002, A\&A, 385, 816
\bibitem[2003]{deb03} de Blok, W.J.G., Bosma, A., McGaugh, S. 2003, MNRAS, 340, 657
\bibitem[2003]{dem03} Demarco, R., Magnard, F., Durret, F., \& M\'arquez, I. 2003, A\&A, 407, 437
\bibitem[2004]{die04} Diemand, J., Moore, B., \& Stadel, J. 2004, MNRAS, 353, 624
\bibitem[2004]{dol04} Dolag, K., Bartelmann, M., Perrotta, F., et al. 2004, A\&A, 416, 853
\bibitem[2003]{ett03} Ettori, S. 2003, MNRAS, 344, L13
\bibitem[2002]{ett02} Ettori, S., De Grandi, S., \& Molendi, S. 2002, A\&A, 391, 841
\bibitem[1995]{fuk95} Fukugita, M., Shimasaku, K., \& Ichikawa, T. 1995, PASP, 107, 945
\bibitem[2004]{gao04} Gao, L., De Lucia, G., White, S.D.M., \& Jenkins, A. 2004, MNRAS, 352, L1
\bibitem[2004]{gav04} Gavazzi, R., Mellier, Y., Fort, B., Cuillandre, J.-C., \& Dantel-Fort, M. 2004, A\&A, 422, 407
\bibitem[2004]{gen04} Gentile, G., Salucci, P., Klein, U., Vergani, D., \& Kalberla, P. 2004, MNRAS, 351, 903
\bibitem[2004]{gil04} Gill, S.P.D., Knebe, A., Gibson, B.K., \& Dopita, M.A. 2004, MNRAS, 351, 410
\bibitem[1996]{gir96} Girardi, M., Fadda, D., Giuricin, G., Mardirossian, F., Mezzetti, M., \& Biviano, A. 1996, ApJ, 457, 61
\bibitem[2005]{gon05} Gonzalez, A.H., Zabludoff, A.I., \& Zaritsky, D. 2005, ApJ, 618, 195
\bibitem[2005]{got05} Goto, T. 2005, MNRAS, 359, 1415
\bibitem[2005]{jee05} Jee, M.J., White, R.L., Ben\'{\i}tez, N., et al. 2005, ApJ, 618, 46
\bibitem[2004]{kat04} Katgert, P., Biviano, A., \& Mazure, A. 2004, ApJ, 600, 657 (KBM)
\bibitem[1998]{kat98} Katgert, P., Mazure, A., den Hartog, R., et al. 1998, A\&AS, 129, 399
\bibitem[1996]{kat96} Katgert, P., Mazure, A., Perea, J., et al. 1996, A\&A, 310, 8
\bibitem[2002]{kel02} Kelson, D.D., Zabludoff, A.I., Williams, K.A., et al. 2002, ApJ, 576, 720
\bibitem[2004]{lin04} Lin, Y.-T., \& Mohr, J.J. 2004, ApJ, 617, 879
\bibitem[2003]{lok03} \L okas, E.L., \& Mamon, G.A. 2003, MNRAS, 343, 401
\bibitem[2006]{lok06} \L okas, E.L., Wojtak, R., Gottl\"ober, S., Mamon, G.A., \& Prada, F. 2006, MNRAS, in press
\bibitem[2006]{mam06} Mamon, G.A., \& Bou\'e, G. 2006, in preparation
\bibitem[1999]{moh99} Mohr, J.J., Mathiesen, B.J., \& Evrard, A.E. 1999, ApJ, 517, 627
\bibitem[1999]{moo99} Moore, B., Quinn, T., Governato, F., Stadel, J., \& Lake, G. 1999, MNRAS, 310, 1147
\bibitem[2004]{mur04} Murante, G., Arnaboldi, M., Gerhard, O., et al. 2004, ApJ, 607, L83
\bibitem[2002]{nat02} Natarajan, P., Loeb, A., Kneib, J.-P., \& Smail, I. 2002, ApJ, 580, L17
\bibitem[1996]{nav96} Navarro, J. F., Frenk, C. S., \& White, S. D. M. 1996, ApJ, 462, 536 (NFW)
\bibitem[2004]{nav04} Navarro, J. F., Hayashi, E., Power, C., et al. 2004, MNRAS, 349, 1039
\bibitem[2002]{nel02} Nelson, A.E., Gonzalez, A.H., Zaritsky, D., \& Dalcanton, J.J. 2002, ApJ, 566, 103
\bibitem[2005]{neu05} Neumann, D.M. 2005, A\&A, 439, 465
\bibitem[2000]{nev00} Nevalainen, J., Markevitch, M., \& Forman, W. 2000, ApJ, 536, 73
\bibitem[2005]{pop05} Popesso, P., Biviano, A., B\"ohringer, H., Romaniello, M., \& Voges, W. 2005, A\&A, 433, 431
\bibitem[2005]{pra05} Pratt, G.W., \& Arnaud, M. 2005, A\&A, 429, 791
\bibitem[2002]{rei02} Reiprich, T.H. \& B\"ohringer, H. 2002, ApJ, 567, 716
\bibitem[2002]{ric03} Ricotti, M. 2003, MNRAS, 344, 1237
\bibitem[2006]{rin06} Rines, K. \& Diaferio, A. 2006, astro-ph/0602032
\bibitem[2003]{rin03} Rines, K., Geller, M.J., Kurtz, M.J., \& Diaferio, A. 2003, AJ, 126, 2152
\bibitem[1999]{sal99} Salucci, P., \& Persic, M. 1999, MNRAS, 309, 923
\bibitem[2004]{san04} Sanchis, T., \L okas, E.L., \& Mamon, G.A. 2004, MNRAS, 347, 1198
\bibitem[2004]{sand04} Sand, D.J., Treu, T., Smith, G.P., \& Ellis, R.S. 2004, ApJ, 604, 88
\bibitem[2006]{sha06} Shankar, F., Lapi, A., Salucci, P., De Zotti, G., \& Danese, L. 2006, ApJ accepted (astro-ph/0601577)
\bibitem[2001]{spr01} Springel, V., White, S.D.M., Tormen, G., \& Kauffmann, G. 2001, MNRAS, 328, 726 
\bibitem[2002]{taka02} Takahashi, K., Sensui, T., Funato, Y., \& Makino, J. 2002, PASJ, 54, 5
\bibitem[2006]{tho06} Thomas, T., \& Katgert, P. 2006, A\&A, 446, 19
\bibitem[2005]{van05} van den Bosch, F.C., Tormen, G., \& Giocoli, C. 2005, MNRAS, 359, 1029
\bibitem[2000]{van00} van der Marel, R., Magorrian, J., Carlberg, R., Yee, H., \& Ellingson, E. 2000, AJ, 119, 2038
\end{thebibliography}
\end{document}